\documentclass[12pt]{article}

\usepackage{graphicx}
\usepackage{times}
\usepackage{natbib}
\usepackage{fancyhdr}

\usepackage{epsfig}
\usepackage{multirow}

\renewcommand{\baselinestretch}{1.5}
\begin{document}

\bibliographystyle{nature}
%


\renewcommand{\multirowsetup}{\centering}

\title{Detecting rich-club ordering in complex networks}

\author{V. Colizza, A. Flammini, M.A. Serrano, A. Vespignani}
\maketitle

\begin{center}
\small{School of Informatics and Department of Physics,
Indiana University, Bloomington 47406 IN}
\end{center}

 %

\noindent {\bf 
Uncovering the hidden regularities and organizational 
principles of networks arising in physical systems ranging 
from the molecular level to the scale of large communication 
infrastructures is the key issue for the understanding of 
their fabric and dynamical 
properties~\cite{Albert:2002,Newman:2003,Doro:2003,Pastorbook,Wasserman:94}. 
The ``rich-club'' phenomenon refers to the tendency of nodes 
with high centrality, the dominant elements of the system, 
to form tightly interconnected communities and it is one of 
the crucial properties accounting for the formation of 
dominant communities in both computer  and social 
sciences~\cite{Pastorbook,Wasserman:94,Price:1986,Zhou:2004,Uzzi:2005}. 
Here we provide the analytical expression and the correct 
null models which allow for a quantitative discussion of 
the rich-club phenomenon. The presented analysis enables 
the measurement of the rich-club ordering and its 
relation with the function and dynamics of networks 
in examples drawn from the biological, social and technological domains.}


Recently, the informatics revolution has made possible the analysis
of a wide range of large scale, rapidly evolving networks such as
transportation, technological, social and biological
networks~\cite{Albert:2002,Newman:2003,Doro:2003,Pastorbook,Wasserman:94}.
While these networks are extremely different from each other in
their function and attributes, the analysis of their fabric provided
evidence of several shared regularities, suggesting
 general and common self-organizing principles beyond
the specific details of the individual systems. In this context, the
statistical physics approach has been exploited as a very convenient
strategy because of its deep connection with statistical graph
theory and because of its power to  quantitatively characterize macroscopic
phenomena in terms of the microscopic dynamics of the various
systems~\cite{Albert:2002,Newman:2003,Doro:2003,Pastorbook,Amaral:2004}.
As an initial discriminant of structural ordering,  attention has
been focused on the networks' degree distribution; i.e., the
probability $P(k)$ that any given node in the network  shares an
edge with $k$ neighboring nodes. This function is, however, only one
of the many statistics characterizing the structural and
hierarchical ordering of a network; a full account of the
connectivity pattern calls for the detailed study of the multi-point
degree correlation functions and/or opportune combination of these.

In this paper, we tackle a main structural property of complex
networks, the so-called ``rich-club'' phenomenon. This property has
been discussed in several instances in both social  
and computer sciences 
and refers to the tendency of high degree nodes, the hubs of 
the network, to be very well connected to each other. 
Essentially, nodes with a large number of links - usually 
referred to as \emph{rich nodes} - are much more likely to
form tight and well interconnected subgraphs (\emph{clubs}) 
 than low degree nodes. A first quantitative definition of the
rich-club phenomenon is given by the rich-club coefficient $\phi$,
introduced by Zhou and Mondragon in the context of the
Internet~\cite{Zhou:2004}. Denoting by $E_{>k}$ the number of
edges among the $N_{>k}$ nodes having degree higher than a given
value $k$, the rich-club coefficient  is expressed as:
\begin{equation}
\phi(k)\,=\,\frac{2 E_{>k}}{N_{>k}(N_{>k}-1)}\,,
\label{eq:phi}
\end{equation}
where $N_{>k}(N_{>k}-1)/2$ represents the maximum possible number of
edges among the $N_{>k}$ nodes. Therefore, $\phi(k)$ measures the
fraction of edges actually connecting those nodes out of the maximum
number of edges they might possibly share. The rich club coefficient
is a novel probe for the topological correlations in a complex
network, and it yields important information about its underlying
architecture. Structural properties, in turn, have immediate
consequences on network's features and tasks, such as e.g.
robustness, performance of biological functions, or selection of
traffic backbones, depending on the system at hand. In a social
context, for example, a strong rich-club phenomenon indicates the
dominance of an ``oligarchy'' of highly connected  and mutually
communicating individuals, as opposed to a structure comprised of many
loosely connected and relatively independent sub-communities. In the
Internet, such a feature would point to an architecture in which
important hubs are much more densely interconnected than peripheral
nodes in order to provide the transit backbone of the
network~\cite{Zhou:2004}. It is also worth stressing that the rich
club phenomenon is not trivially related to the mixing properties of
networks, which enable the distinction between assortative
networks, where large degree nodes preferentially attach to large
degree nodes, and disassortative networks, showing the opposite
tendency~\cite{Pastor:2001,Newman:2002,Maslov:2002}. Indeed, the
rich club phenomenon is not necessarily associated to assortative
mixing. In the top panel of Fig. 1, we sketch a
simple construction in which a disassortative network is exhibiting
 the rich club phenomenon. In other words, the rich
club phenomenon and the mixing properties express different features
that are not trivially related or derived one from each other (the
technical discussion of this point is reported in the methods
section).

In Fig. 1, we report the behavior of the rich club
coefficient as a function of the degree in a variety of real world
networks drawn from the biological, social and technological world.
In Table 1, we summarize the basic topological
features of these networks and the datasets used. We also consider
three standard network models: the Erd\"os-R\'enyi (ER)
graph~\cite{ER}, the generalized random network having a
heavy-tailed degree distribution obtained with the Molloy-Reed (MR)
algorithm~\cite{MR}, and the Barabasi-Albert (BA) model~\cite{BA}.
In the ER graph, $N$ nodes are connected by $E$ edges randomly
chosen with probability $p$ out of the $N(N-1)/2$ possible pairs of
nodes. The MR network is  obtained starting from a given
degree sequence $P(k)$ (in our case $P(k)\sim k^{-\gamma}$ with
$\gamma=3$) by randomly connecting nodes with the constraints of
avoiding self-loops and multiple edges. 
 The BA model is generated by using the growing
algorithm of Ref.~\cite{BA} that produces a scale-free graph with
power-law degree sequence with exponent $\gamma=3$. 
 In all cases, the generated
networks have $N=10^5$ vertices and an average degree $\langle
k\rangle=6$.

As is evident from Fig. 1, the monotonic increasing
 of $\phi(k)$ is a feature shared by all the analyzed
datasets. This behavior is claimed to provide evidence of the
rich-club phenomenon since $\phi(k)$  progressively increases in
vertices with increasing degree ({\it e.g.}, see
Ref.~\cite{Zhou:2004} for the Internet case, where a different
representation of the function is adopted with $\phi$ defined in
terms of the rank $r$ of nodes sorted by decreasing degree values).
However, a monotonic increase of $\phi(k)$ does not necessarily
implies the presence of the rich-club phenomenon. Indeed, even in
the case of the ER graph - a completely random network - we find an
increasing rich-club coefficient. This implies that the increase of
$\phi(k)$ is a natural consequence of the fact that vertices with
large degree have a larger probability of sharing edges than low
degree vertices. This feature is therefore imposed by construction
and does not represent a signature of any particular organizing
principle or structure, as  is clear in the ER case.
 The simple inspection of the $\phi(k)$ trend is therefore
potentially misleading in the discrimination of the rich-club
phenomenon.

In order to find  opportune baselines for the detection of the
rich-club phenomenon we focus on the theoretical analysis of
$\phi(k)$. In the methods section we derive an expression for the
rich club coefficient as a function of the convolution of the two
vertices degree correlation function $P(k,k')$. Interestingly, it is
possible to obtain an explicit expression for the rich-club
coefficient of random uncorrelated networks. In this case, the two-vertices
correlation function is a simple function of the degree
distribution, yielding the following behavior for uncorrelated large
size networks at large degrees:
\begin{equation}
\phi_{unc}(k)
\begin{array}{*{1}c} \\ \sim \\ ^{k, k_{max} \rightarrow \infty}
\end{array}\frac{k^2}{\langle k
\rangle N}\,\, , \label{eq:phiunck}
\end{equation}
where $k_{max}$ is the maximum degree present in the
network. Eq.(\ref{eq:phiunck}) shows unequivocally that the rich-club
coefficient is also a monotonically increasing function  for
uncorrelated networks, so that, in order to assess the presence of
rich-club structural ordering,  it is necessary to compare it with
the one obtained from the appropriate null model with the same degree 
distribution, thus providing a suitable
normalization of $\phi(k)$.

From the previous discussion, a possible choice for the
normalization of the rich-club coefficient is provided by the ratio
$\rho_{unc}(k)=\phi(k)/\phi_{unc}(k)$, where $\phi_{unc}(k)$ is
analytically calculated by inserting in Eq.~(\ref{eq:phiuncM}),
reported in the methods section, the network's degree distribution
$P(k)$. A ratio larger than one is the actual evidence for the
presence of a rich-club phenomenon leading to an increase in the
interconnectivity of large degree nodes in a more pronounced way
than in the random case. On the contrary, a ratio $\rho_{unc}(k)<1$
is a signature of an opposite organizing principle that leads to a
lack of interconnectivity among large degree nodes. 
On the other hand, a completely degree-degree
uncorrelated network with finite size is not always realizable due
to structural constraints. Indeed, any finite size random network
presents a structural cut-off value $k_s$ over which the requirement
of the lack of dangling edges introduces  the presence of multiple
and self-connections and/or degree-degree
correlations~\cite{Boguna:2004,moreira:2002}. Networks with bounded
degree distributions and finite second moment $\langle k^2\rangle$
present a $k_{max}$ that is below the structural one $k_s$. In this
situation, $\phi_{unc}(k)$ is properly defined for all degrees and
is representative of the network topology. However, in networks with
heavy-tailed degree distribution (e.g., scale-free degree
distributions with $2 < \gamma \leq 3$, as observed in many real
systems), this is no longer the case and $k_s$ is generally smaller
than $k_{max}$. In fact, structural degree-degree correlations
and higher order effects, such as the emergence of large
cliques~\cite{Bianconi:2005}, set in even in completely random
networks. The normalization of $\phi(k)$ that takes into account 
these effects is provided by the expression
$\rho_{ran}(k)=\phi(k)/\phi_{ran}(k)$, where $\phi_{ran}(k)$ is the
rich-club coefficient of the maximally random network with the same degree
distribution $P(k)$ of the network under study.
Operatively, the maximally random network can be thought of as the
stationary ensemble of networks visited by a process that, at any
time step, randomly selects a couple of links of the original
network and exchange two of their ending points (automatically
preserving the degree distribution). Also in this case an actual
rich-club ordering is denoted by a ratio $\rho_{ran}(k)>1$.
Therefore, whereas $\rho_{unc}(k)$ provides information about the 
overall rich-club ordering in the network with respect to an 
ideally uncorrelated
graph, $\rho_{ran}(k)$ is a normalized measure which discounts the
structural correlations due to unavoidable finite size effects,
providing a better discrimination of the actual presence of 
the rich club-phenomenon 
due to the ordering principles shaping the network.

In Fig. 2, we report the ratios $\rho_{ran}(k)$ 
for the real world and the simulated networks. The
analysis clearly discriminates between networks with or without
rich-club ordering. In particular, we identify a strong rich-club
ordering in the Scientific Collaboration Network, providing support
to the idea that the elite formed by more influential scientists
tends to form collaborative groups within specific domains. This
also supports the view that the rich-club phenomenon is a natural
tendency in many social networks. We find a clearly opposite result
in the decreasing behavior of the rich club spectrum for the Protein
Interaction Network and the Internet map at the Autonomous
System level. In both
cases, this evidence provides interesting information regarding the system
structure and function. 

The lack of rich-club ordering in the
Protein Interaction Network indicates that proteins with  large
number of interactions are presiding over different functions and
thus, in general, are  coordinating specific functional modules
(whose detailed analysis requires specific tools~\cite{guimera:2005}).
Figure 3 shows portions of the Protein Interaction
Network  and the Scientific Collaboration Network
including the club of $N_{>k}$ nodes -- $N_{>k}=29$ and $N_{>k}=35$ for the 
Protein Interations,
$N_{>k}=30$ and $N_{>k}=36$ for the Scientific Collaboration  -- and the
connections among them. The
network representations clearly show the presence of a rich-club
phenomenon in the Scientific Collaboration Network, where the
majority of rich nodes are highly interconnected forming tight
subgraphs, in contrast with the Protein Interaction Network case,
where only few links appear to connect rich nodes, the rest linking
to lower degree vertices.

In the case of the Internet, the appropriate analysis of the
rich-club phenomenon shows that, contrary to previous
claims~\cite{Zhou:2004}, the structure at the Autonomous
System level {\em lacks} rich-club
ordering. This might appear counter-intuitive. It is reasonable to
imagine the Internet backbone made of interconnected transit
providers which are also local hubs. This is however not the case
and an explanation can be easily found in the fact that we are just
considering topological properties. Indeed, the backbone hubs are
identified more in terms of their bandwidth and traffic capacity
than in terms of the sole number of connections. The present result
suggests that high degree hubs provide connectivity to local region
of the Internet and are not tightly interconnected. The backbone of
interconnected transit providers is instead identified by high
traffic links which play a crucial role in terms of traffic
capacities but whose number might represent a small fraction of the
total possible number of interconnections.

The previous discussion points out that, in some cases, the concept
of rich-club ordering should be generalized in order to evaluate the
richness of vertices not just in terms of their degree but in terms
of the actual traffic or intensity of interactions handled. In this
case, we have to consider a weighted network representation of
the system where a weight $w_{ij}$ representing the traffic or
intensity of interaction is associated to each edge between the
vertices $i$ and $j$. Also in this case, however, the study of
the weighted rich-club coefficient alone does not discriminate the
actual presence of the rich club effect (see Methods). 
Given the entanglement of the weight and
degree correlations, the appropriate null hypothesis is however
more complicated to define and a detailed account of the
evaluation of the weighted rich-club effect will be provided
elsewhere.

In summary, the presented analysis provides the baseline functions for the
detection of the rich-club phenomenon and its effect on the
structure of large scale networks. This allows the measurement of
this effect in a wide range of systems, finally enabling a
quantitative discussion of various claims such as ``high
centrality'' backbones in technological networks and ``elitarian''
clubs in social systems.

\section*{Methods}
{\bf Analytic expression of the rich club coefficient}. The basic
analytical understanding of the rich-club phenomenon starts by
considering the quantity $E_{kk'}$, representing the total number of
edges between vertices of degree $k$ and of degree $k'$ for $k\neq
k'$, and twice the number of edges between vertices in the same
degree class. We can express the numerator of $\phi(k)$ in
Eq.~\ref{eq:phi} as $2E_{>k}\,=\int_k^{k_{max}}dk'\int_k^{k_{max}}
dk''E_{k'k''}$, where $k_{max}$ is the maximum degree present in the
network and where, for the sake of simplicity, the variable $k$ is
thought of as continuous. In turn, the quantity $E_{kk'}$ can be
expressed as a function of the joint degree probability
distribution~\cite{Newman:2002,Boguna:2004,Boguna:2003,newman:2003}
via the identity $N\langle k\rangle P(k,k')=E_{kk'}$, yielding
\begin{equation}
\phi(k)\,=\, \frac{N\langle k\rangle\int_k^{k_{max}}dk'\int_k^{k_{max}}dk''
P(k',k'')}{\left[ N\int_k^{k_{max}}dk' P(k') \right]\left[ N\int_k^{k_{max}}dk' P(k')-1 \right] }.
\label{eq:phi_def}
\end{equation}
From Eq.~(\ref{eq:phi_def}), it is clear that $\phi(k)$ is also a
measure of correlations in the network, although it represents a
different projection of $P(k,k')$ as compared to other degree-degree
correlation measures. At the same time, it is  possible to see
that the rich-club coefficient express a property that is not
trivially related to the usual indicators of assortative behavior,
such as the Pearson's correlation coefficient~\cite{Newman:2002} or
the average nearest neighbor degree~\cite{Pastor:2001}. Notice that
these assortativity measures quantify two-point correlations and so
account for quasi-local properties of the nodes in the network,
whereas the rich club phenomenon is computed as a global feature
within a restricted subset. The double integral is indeed a
convolution of the correlation function that allows the presence of different
combinations of the assortative and rich-club features in the same network.

Only in the case of random uncorrelated
networks~\cite{Doro:2003,Pastorbook,newman:2003}, the joint degree
distribution $P(k,k')$ factorizes and takes the simple form
$P_{unc}(k,k')=kk'P(k)P(k')/\langle k\rangle^2.$ By inserting this
expression into Eq.~(\ref{eq:phi_def}), we obtain $\phi(k)$ for
uncorrelated networks as
\begin{equation}
\phi_{unc}(k)=\frac{1}{N\langle k\rangle} \left[ \frac{\int_k^{k_{max}}
dk' k' P(k')}{\int_k^{k_{max}} dk' P(k')}  \right]^2
\begin{array}{*{1}c} \\ \sim \\ ^{k, k_{max} \rightarrow \infty}
\end{array}\frac{k^2}{\langle k
\rangle N}\,\, , \label{eq:phiuncM}
\end{equation}
where we have applied L'H\^{o}pital's rule to derive the behavior
for large size networks and large degrees.

{\bf Rich club coefficient for weighted networks}.
If the rich-club is defined as the set of
nodes having a strength  larger than a given value $s$, a possible
definition of the \emph{weighted rich-club coefficient} can be
expressed as
\begin{equation}
\phi^w(s)\,=\,\frac{2W_{>s}}{\sum_{i|s_i>s} s_i}\,,
\end{equation}
where $W_{>s}$ represents the sum of the weights on the links connecting
two nodes in the club and the normalization is given by the sum of the
strengths of the rich nodes.


\bigskip
\noindent {\bf Acknowledgments}

We thank M. Bogu\~n\'{a}, M. Barth{\'e}lemy, S. Wasserman and E. Flach
for useful discussions and suggestions. A.V. is partially supported by the NSF
award IIS-0513650.

Correspondence and requests for materials should be addressed to
A.V.

\newpage
\section*{Table legend}

\noindent
{\bf Table 1: Basic topological properties of the
analyzed datasets.}
We considered four real world networks: (1) the
{\em Protein Interaction Network}~\cite{Maslov:2002,Colizza:2005}
 of the yeast \emph{Saccharomyces Cerevisiae} collected with
different experimental techniques and documented at the Database of
Interacting Proteins (http://dip.doe-mbi.ucla.edu/); (2) the {\em
Scientific Collaboration Network}~\cite{Newman:2001_1,Newman:2001_2}
 extracted from the electronic database e-Print Archive in the
area of condensed matter physics
(http://xxx.lanl.gov/archive/cond-mat/), from 1995 to 1998, in which
nodes represent scientists and a connection exists if they
coauthored at least one paper in the archive; (3) the network of {\em Worldwide
Air Transportation }~\cite{Barrat:2004,Guimera:2005}  representing
the International Air Transport Association (http://www.iata.org/)
database of airport pairs connected by direct flights for the year
2002; (4) the {\em Internet} network at the Autonomous
System~\cite{Pastorbook}
level~\cite{Pastorbook,Pastor:2001,Faloutsos:1999,Vazquez:2002,Quian:2002}
 from data collected by the {\it Oregon Route Views} project
(http://www.routeviews.org/) in May 2001, in which nodes represent
Internet service providers and edges connections among those. The
sizes of the networks in number of nodes and edges are shown, along
with the average degree $\langle k\rangle$ and the maximum degree
value $k_{max}$. We also give the value for the corresponding
structural cut-off, $k_s$, in the uncorrelated
case~\cite{Boguna:2004}.

\section*{Figure Legends}

\noindent
{\bf Figure 1: Schematic picture of the rich-club phenomenon and
rich-club spectrum $\phi(k)$ for real networks.} 
At the top, a conceptual example of disassortative
network  displaying the presence of the rich-club phenomenon is shown.
Disassortative mixing is given by the tendency of hubs to be on average
more likely connected to low degree nodes. However, the four rich nodes
represented in the schematic picture show a clear rich-club behavior
by forming a fully connected clique within the club.
At the bottom, results for the four
real-world networks and the three models analyzed are shown. The
computer generated networks - ER, MR, and BA - have size $N=10^5$ and
average degree $\langle k\rangle=6$. ER refers to the
\emph{Erd\"os-R\'enyi graph}, MR is constructed from the
\emph{Molloy-Reed} algorithm with a given degree distribution
$P(k)\sim k^{-3}$, and the BA model is generated by growing a
network with preferential attachment that produces a scale-free
graph with power-law degree sequence with exponent $\gamma=3$.
Results are averaged over $n=10^2$ different realizations for each
model. All networks share a monotonic increasing behavior of
$\phi(k)$, independent of the nature of the degree distribution
characterizing the network and of the possible presence of
underlying structural organization principles. Also random networks,
either having a Poissonian degree distribution (such as ER) 
or a heavy-tailed $P(k)$ (such as MR and BA), 
show a rich club spectrum increasing with increasing
values of the degree. This common trend is indeed due to an
intrinsic feature of every network structure, for which hubs have
simply a larger probability of being more interconnected than low
degree nodes. 

\bigskip
\noindent
{\bf Figure 2: Assessment for the presence of the rich-club
phenomenon in the networks under study.}
$\phi(k)$ is compared to
the null hypothesis provided by the maximally
random network with $\phi_{ran}(k)$. The ratio
$\rho_{ran}=\phi/\phi_{ran}$ is
plotted as a function of the degree $k$ and compared  to the baseline
value equal to $1$. If $\rho(k)>1\,(<1)$ the network displays the
presence (absence) of the rich-club phenomenon with respect to the
random case. The Protein Interaction Network, the Internet map
at the Autonomous System level  and the Scientific Collaboration
Network  show clear behaviors as explained in the main text.
The Worldwide Air Transportation network  displays a mild
rich-club ordering with $\rho_{ran}(k)>1$. The ER and MR
network models show a ratio $\rho_{ran}(k)=1\,\forall k$, as
expected, whereas the BA model exhibits a mixing behavior with
values above $1$ for very high degrees.

\bigskip
\noindent
{\bf Figure 3: Graph representations of the rich-clubs.}
Progressively smaller 
clubs of
$N_{>k}$ rich nodes in the Protein Interaction Network -top-
and in the Scientific Collaboration Network  -bottom- are shown
together with the $E_{>k}$ connections among them. Here
$N_{>k}=35,\,E_{>k}=37$ (top left) and $N_{>k}=29,\,E_{>k}=21$ (top right)
for the Protein Interactions; $N_{k>}=36,\,E_{>k}=62$ (bottom left) and 
$N_{k>}=30,\,E_{>k}=54$ (bottom right) for the 
collaboration network. The two graph representations for each network show progressively smaller
clubs made of  $N_{>k}$ rich nodes for increasing values of the degree $k$.
The links connecting the rich nodes to the
rest of the network are not represented for sake of simplicity. 
The Protein Interaction Network shows a club whose
hubs are relatively independent being loosely connected among each
other, leaving the remaining links to coordinate specific functional
modules. A different picture is observed in the Scientific Collaborations
 case, where most
of the hubs form cliques and tightly interconnected subgraphs, thus
revealing the tendency of scientists to form densely interconnected
collaborative groups. The graphs have been produced with the Pajek
software (http://vlado.fmf.uni-lj.si/pub/networks/pajek/).

 \newpage
\begin{table}[!ht]
\begin{center}
\vspace{2cm}
\begin{tabular}{|l||c|c|c|c|}
\hline
\vspace{-.3cm}
 & ~{\bf Protein}~ & ~{\bf Scientific}~ & ~{\bf Air}~ & ~{\bf Internet}~  \\
 & ~{\bf Interactions}~ & ~{\bf Collaborations}~ & ~{\bf Transportation}~ & \\
\hline
\hline
\# nodes~ & 4713 & 15179 & 3880 & 11174  \\
\hline
\# links & ~14846~ & ~43011~ & ~18810~ & 23409\\
\hline
$\langle k\rangle$ & 6.3 & 5.7 & 9.7 & 4.2 \\
\hline
$k_{max}$ & 282 & 97 & 318 & 2389 \\
\hline \hline
$k_{s}=\sqrt{\langle k\rangle N}$ & 172 & 294 &  194 &  216 \\
\hline
\end{tabular}
\vspace{1cm} 
\renewcommand{\baselinestretch}{1.0}
\label{tab:features}

{\bf Table 1.}
\end{center}
\end{table}

\newpage
\begin{figure}[!ht]
\begin{center}
\vskip .7cm
\includegraphics[width=16cm]{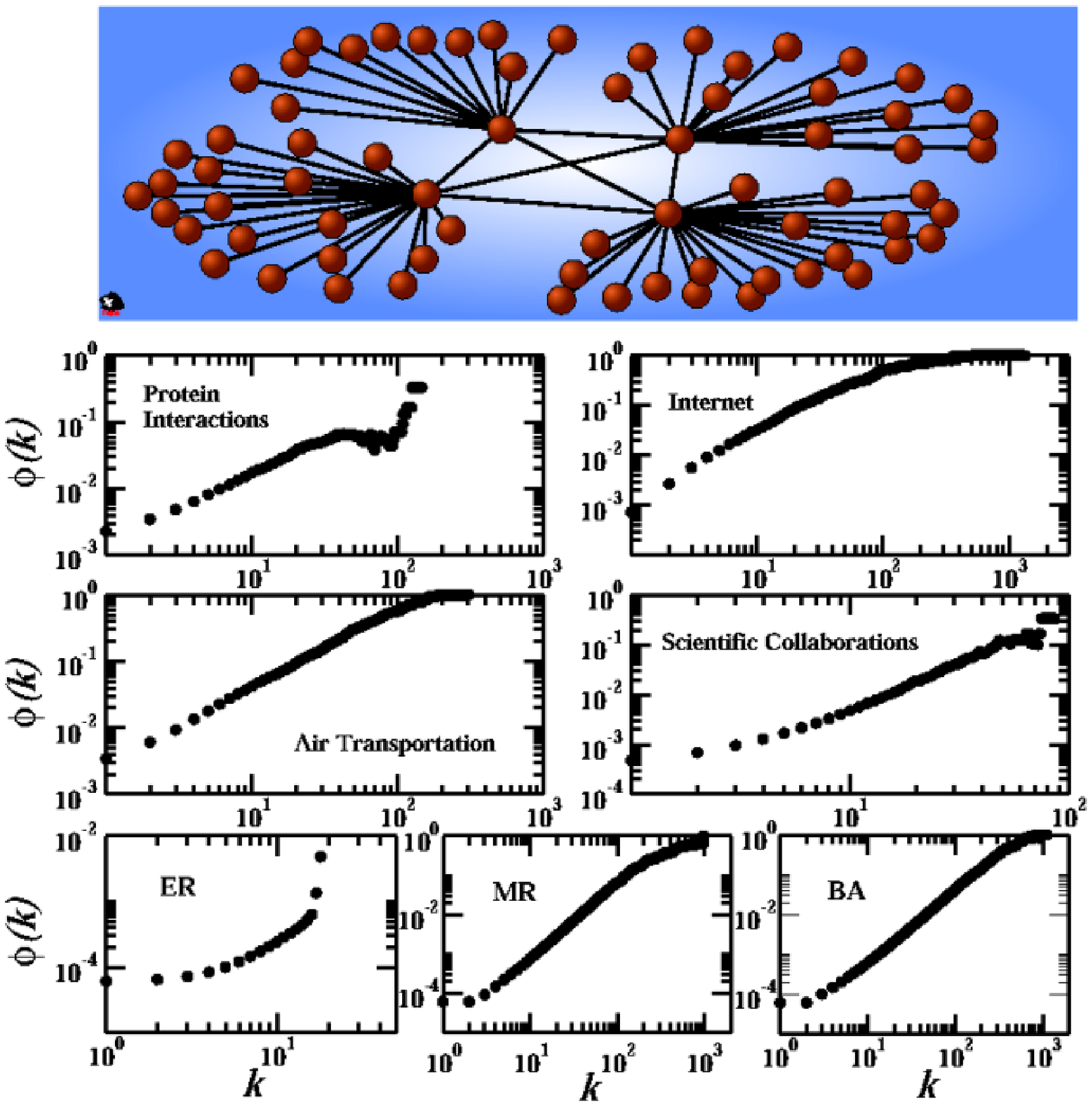}
\vspace{.3cm}
\label{fig:phi}

{\bf Figure 1.}
\end{center}
\end{figure}

\newpage
\begin{figure}[!ht]
\begin{center}
\vskip 2cm
\includegraphics[height=16cm,angle=-90]{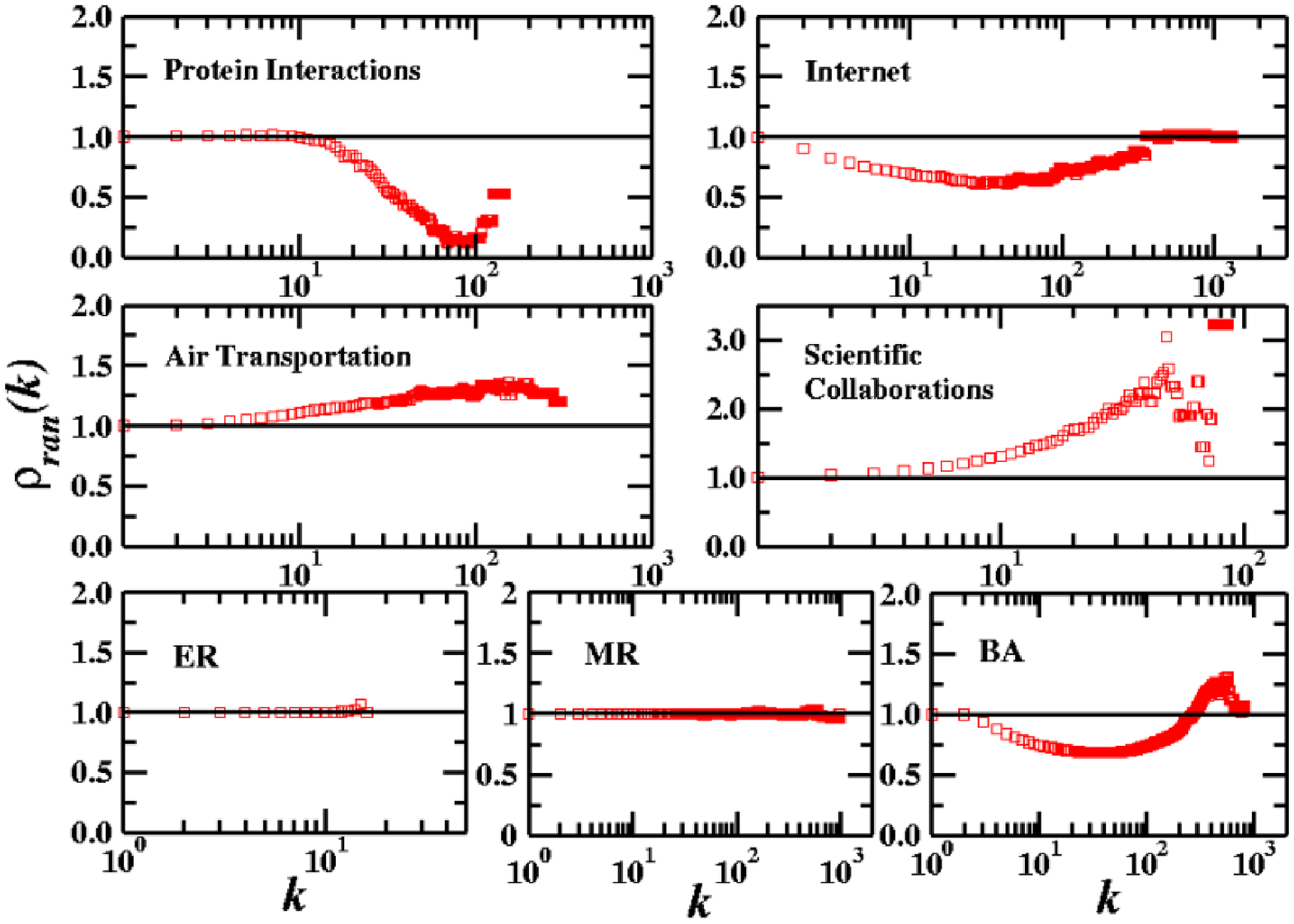}
\vskip .1cm
\renewcommand{\baselinestretch}{1.0}
\label{fig:phi_ratio}
\vspace{.3cm}

{\bf Figure 2.}
\end{center}
\end{figure}

\newpage
\begin{figure}[!ht]
\begin{center}
\includegraphics[width=16cm]{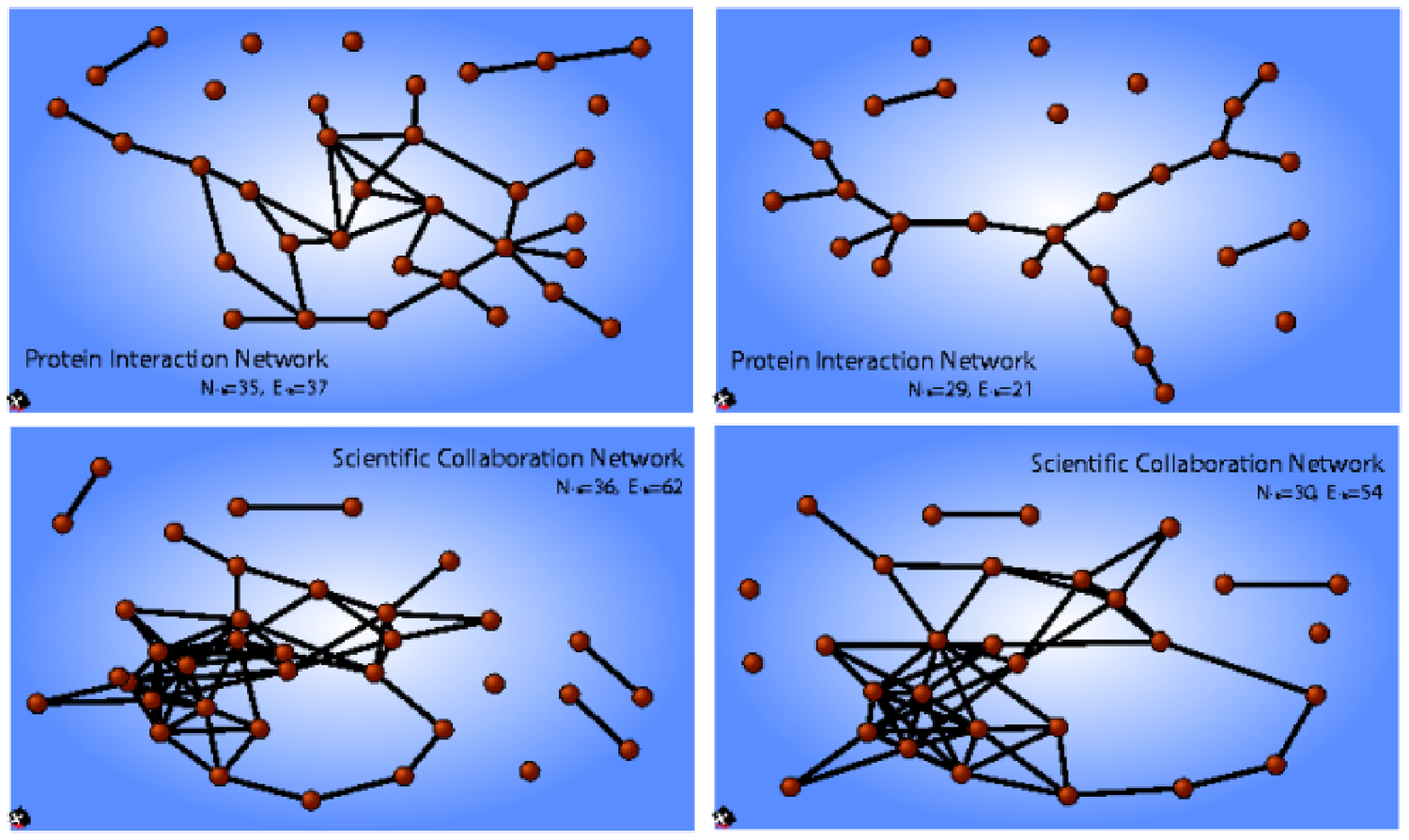}
\renewcommand{\baselinestretch}{1.0}
\label{fig:networks}
\vspace{.3cm}

{\bf Figure 3.}
\end{center}
\end{figure}


\end{document}